\documentclass[a4paper,12pt]{article}
\begin{document}
\vskip 1.8cm
\centerline{\Large \bf ${\cal{N}}=1$ super Yang-Mills from D branes 
\footnote{Work partially supported by the European
Commission RTN programme HPRN-CT-2000-00131.}} 
\vskip 1.4cm \centerline{\bf P. Di Vecchia} \vskip .8cm
\centerline{\sl $^a$ NORDITA, Blegdamsvej 17, DK-2100 Copenhagen
\O, Denmark} \centerline{\tt divecchi
@alf.nbi.dk} 
\begin{abstract}
We use fractional and wrapped branes to describe perturbative and
non-perturbative properties of ${\cal{N}}=1$ super Yang-Mills living on their
world-volume. 
\end{abstract}
\renewcommand{\thefootnote}{\arabic{footnote}}
\setcounter{footnote}{0} \setcounter{page}{1}

\section{Introduction}

String theory emerged in the beginning of the seventies from the
attempt to find a theory describing strong interactions~\cite{GSW}. 
It was soon
realized that it contained many unphysical features for describing
strong interactions as for instance 
the existence of massless spin $1$ and $2$ particles in the hadronic
spectrum and extra
dimensions. For these reasons it was later proposed~\cite{SS} that string
theories were more suited to describe a unifying theory of all
interactions  and a description of
strong interactions based on some kind of string theory is still to be
found. In the meantime it became clear that strong interactions are
described by QCD that is a nonabelian gauge theory based on the gauge
group $SU(3)$. 

The discovery of D branes as non-perturbative states of string theory
has open again the way to use string theory for describing the
properties of the gauge theories that live on their world-volume. This
has been possible because the D branes have the twofold property of
being, on the one hand, a classical solution of the low-energy string
effective action containing closed string states and of containing, on 
the other hand, a gauge theory living on its world-volume whose
degrees of freedom correspond to open strings having their end-points 
attached to the world-volume of the D brane. This is a direct consequence of
open/closed string duality according to which open-string loop
diagrams can equivalently be described by tree diagrams of closed
strings. The most impressive consequence of this duality has been the
Maldacena conjecture~\cite{MALDA} that implies the exact equivalence between
four-dimensional ${\cal{N}}=4$ super Yang-Mills and type IIB string
theory compactified on $AdS_5 \times S^5$.   

Recently these ideas have been extended to more realistic gauge
theories that are less supersymmetric and non-conformal and although
in this case
no exact duality has been established, nevertheless 
they have allowed
to derive perturbative and non-perturbative properties of the gauge
theories living on the world-volume of D branes from their
supergravity description~\footnote{For  extensive reviews of
these developments see Refs.~\cite{KLEBARE,MATTEO,MIL,HOUCHE,EMILI}.} 
Those more realistic gauge theories can be
obtained by considering more sophisticated D branes as fractional D branes
of some orbifold that are stack at the orbifold fixed point or as
branes wrapped on some nontrivial two-cycle of a Calabi-Yau space.

In this talk I want  to discuss the results obtained restricting
myself to ${\cal{N}}=1$ super Yang-Mills where also
non-perturbative properties as the gaugino condensantion and the
non-perturbative effective potentials  have been derived. 
It turns out that, in
order to derive non-perturbative properties, we need to consider 
supergravity solutions that are regular also at short distances. 
Two of them are known, namely the one
corresponding to a D5 brane wrapped on a notrivial two-cycle of a
Calabi-Yau space~\cite{MN} and the one corresponding
to the deformed conifold found in Ref.~\cite{KS}. 
In the next section we will describe them and from them we will derive
the properties of ${\cal{N}}=1$ super Yang-Mills.

An important property of the gauge theories that can be derived with
these methods are the running coupling constant that is given by the
following general formula:
\begin{equation}
\frac{4 \pi}{g_{YM}^{2}} =  \frac{1}{g_s ( 2 \pi \sqrt{\alpha'})^{2}} \int
d^{2} \xi {e}^{- (\phi - \phi_0)} \sqrt{\det{( G_{AB} + B_{AB})}}
\label{volu87}
\end{equation}
and the $\theta$ parameter given by:
\begin{equation}
\theta_{YM} = 
\frac{1}{2 \pi \alpha' g_s}
\int_{{\cal{C}}_2} ( C_2 + C_0  B_2)
\label{theta56}
\end{equation}
The previous formulas are valid for both fractional and wrapped branes 
as it can be found in Ref.s~\cite{MATTEO,HOUCHE,EMILI}.

\section{${\cal{N}}=1$ super Yang-Mills from D branes}

In this section we will review the properties of the two regular
solutions of ten-dimensional type IIB string theory discussed in the 
introduction and we will extract from them
the properties of ${\cal{N}}=1$ super Yang-Mills.

\subsection{The MN solution}

We start by writing the classical solution corresponding to $N$ D branes
wrapped on a two-cycle of a non-compact Calabi-Yau space found in 
Ref.~\cite{MN}. It has a non-trivial metric:
\begin{equation}
d s_{10}^{2}
={\rm e}^\Phi
\left[  dx_{1,3}^{2} +
  \frac{{\rm e}^{2h}}{\lambda^2} \left(d {\widetilde{\theta}}^2
+ \sin^2 {\widetilde{\theta}}\,d {\widetilde{\varphi}}^2 \right)
\right]
+\frac{{\rm e}^\Phi
}{\lambda^2} \left[ d \rho^2 +
\sum_{a=1}^3\left( \sigma^a - \lambda A^a \right)^2\right]~~,
\label{stri987}
\end{equation}
a two-form R-R potential 
\begin{eqnarray}
{C}^{(2)} &=& \frac{1}{4 \lambda^2} \left[ (\psi + \psi_0 ) \left(\sin
\theta'\, d \theta'
  \wedge d \phi - \sin {\widetilde{\theta}} \,d {\widetilde{\theta}}
  \wedge d {\widetilde{\varphi}}
\right) - \cos \theta' \,\cos
  {\widetilde{\theta}} \,
d \phi \wedge d {\widetilde{\varphi}}\right]\nonumber \\
&&+\,\frac{a}{2\lambda^2} \left[ d{\widetilde{\theta}} \wedge
\sigma^1 - \sin {\widetilde{\theta}}\,
 d{\widetilde{\varphi}} \wedge \sigma^2 \right] 
\label{f3exact}
\end{eqnarray}
and a dilaton
\begin{equation}
{\rm e}^{2 \Phi} =
\frac{\sinh 2 \rho}{2 \,{\rm e}^h}~~,
\label{c298}
\end{equation}
where
\begin{equation}
{\rm e}^{2h} =  \rho \coth 2 \rho - \frac{\rho^2}{\sinh^2 2
\rho} - \frac{1}{4}~~,~~ 
{\rm e}^{2k} =  {\rm e}^{h}\,\frac{\sinh 2\rho}{2}~~,~~ 
a = \frac{2 \rho}{\sinh 2 \rho} 
\label{equa}
\end{equation}
and
\begin{equation}
A^{1} = -\,\frac{1}{2\lambda}\, a(r)\, d{\widetilde{\theta}}~~,~~
A^{2} =  \frac{1}{2\lambda}\,a(r) \sin {\widetilde{\theta}}\, d
{\widetilde{\varphi}}~~,~~ A^{3} = -\,\frac{1}{2\lambda}\,\cos
{\widetilde{\theta}}\, d
 {\widetilde{\varphi}}~~.
\label{vec45}
\end{equation}
with $\rho \equiv \lambda \,r$ and $\lambda^{-2}= N g_s \alpha'$. 
The left-invariant 1-forms of
$S^3$ are
\[
\sigma^1 =  \frac{1}{2}\Big[\cos \psi \,d\theta' + \sin \theta'
\sin \psi \,d \phi  \Big]~~,~~\sigma^2 = -\frac{1}{2} \Big[\sin
\psi \,d \theta' - \sin \theta' \cos \psi \,d\phi \Big]~~,
\]
\begin{equation}
\sigma^3 = \frac{1}{2} \Big[d \psi + \cos \theta' \,d \phi
\Big]~~,
\label{diffe49}
\end{equation}
with $0\leq \theta'\leq\pi$, $0\leq\phi\leq 2\pi$ and
$0\leq\psi\leq 4\pi$. The variables ${\tilde{\theta}}$ and
${\tilde{\varphi}}$ describe a two-dimensional sphere and vary in the
range $0 \leq {\tilde{\theta}} \leq \pi$ and 
$ 0 \leq {\tilde{\varphi}} \leq 2 \pi$.   

We can now use the previous solution for computing the running
coupling constant and the $\theta$ parameter of ${\cal{N}}=1$ super
Yang-Mills~\footnote{See Ref.s~\cite{MATTEO,HOUCHE,EMILI} and Ref.s therein.}. 
In order to do that we have to fix the cycle around which
we should perform the integrals in eq.s (\ref{volu87}) and
(\ref{theta56}). It turns out that this two-cycle is specified by:
\begin{equation}
{\tilde{\theta}} =  \theta' ~~.~~{\tilde{\varphi}}= - \phi~~,~~ \psi=0
\label{cycle2}
\end{equation}
keeping  $\rho$ fixed. If we now compute the gauge
couplings on the cycle specified in the previous equation ($B_2 = C_0 =0$)
we get
\begin{equation}
\frac{4 \pi^2}{N g^{2}_{YM}} = \rho \coth 2  \rho 
+ \frac{1}{2} a (\rho) \cos \psi
\label{gaucou98}
\end{equation}
and
\begin{equation}
\theta_{YM} = \frac{1}{2 \pi g_s \alpha'}  \int_{S^2} C_2 =
 - N \left( \psi + a (\rho) \sin \psi  + \psi_0 \right)
\label{theta45}
\end{equation}
where  we have not yet taken $\psi =0$ for reasons
that will become clear in a moment. Eq.~ (\ref{gaucou98}) shows that
the coupling constant is running as a function of the distance $\rho$ from
the branes. In order to obtain the correct running of the gauge
theory we have to find a relation between $\rho$ and the
renormalization group scale $\mu$. This can be obtained with the following
considerations. 
If we look at the Maldacena-N{\'{u}}\~nez solution it is easy
to see that the metric in Eq.(\ref{stri987}) is invariant under
the following transformations:
\begin{equation}
\left\{ \begin{array}{c}
                \psi \rightarrow \psi + 2 \pi~~~~~if~~ a \neq 0 \\
                \psi \rightarrow  \psi + 2 \epsilon~~~~~if~~ a =0
\end{array} \right.
\label{arr12}
\end{equation}
where $\epsilon$ is an arbitrary constant. On the other hand $C_2$ is
not invariant under the previous transformations but its flux, that is
exactly equal to $\theta_{YM}$ in Eq.(\ref{theta45}), changes
by an integer multiple of $2 \pi$. In fact one gets:
\begin{equation}
\theta_{YM} = \frac{1}{2 \pi \alpha' g_s} \int_{{\cal{C}}_2}  C_2 \rightarrow
\theta_{YM} +
\left\{ \begin{array}{c}
             - 2 \pi N~~,~~~ if~~ a \neq 0\\
             - 2 N \epsilon~~,~~if~~ a=0,\epsilon= \frac{\pi k}{N}
\end{array} \right.
\label{arr2}
\end{equation}
This changes $\theta_{YM}$ by a factor $2 \pi$ times an integer.
But since the physics does not change when 
$ \theta_{YM} \rightarrow \theta_{YM} + 2 \pi$ this
implies that the transformation in Eq.(\ref{arr2}) is an invariance. Notice
that also Eq.(\ref{gaucou98}) for the gauge coupling constant, is
invariant under the transformation in Eq.(\ref{arr12}). This means
that the classical solution and also the gauge couplings  are
invariant under the $Z_2$ transformation if $a \neq0$, while this
symmetry becomes $Z_{2N}$ if $ a$ is taken to be zero. This
implies that, since  in the ultraviolet $a (\rho)$ is
exponentially small, we can neglect it and we have a $Z_{2N}$
symmetry, while in the infrared  we cannot neglect $a (\rho)$
anymore and we have only a $Z_2$ symmetry left. It is on the other
hand well known that ${\cal{N}}=1$ super Yang-Mills has a non zero
gaugino condensate $<\lambda \lambda>$ that is responsible for the
breaking of $Z_{2N}$ into $Z_2$. Therefore it is natural to
identify the gaugino condensate with the function $ a(\rho)$ that
appears in the supergravity solution:
\begin{equation}
< \lambda \lambda> \sim \Lambda^3 = \mu^3 a( \rho )
\label{gauar78}
\end{equation}
This  gives the relation between the renormalization group scale $\mu$
and the supergravity space-time parameter $\rho$.
In the ultraviolet (large $\rho$) $a(\rho)$ is exponentially
suppressed and in Eq.s (\ref{gaucou98}) and (\ref{theta45})
we can neglect it obtaining:
\begin{equation}
\frac{4 \pi^2}{N g^{2}_{YM}} = \rho \coth 2 \rho ~~,~~
\theta_{YM} = - N \left( \psi  + \psi_0 \right)
\label{gautheta}
\end{equation}
The chiral anomaly can be obtained by performing the transformation
$\psi \rightarrow \psi + 2 \epsilon$ and getting:
\begin{equation}
\theta_{YM} \rightarrow \theta_{YM} - 2N \epsilon
\label{chi67}
\end{equation}
This implies that the $Z_{2N}$ transformations corresponding to
$\epsilon = \frac{\pi k}{N}$ are symmetries because they shift
$\theta_{YM}$ by multiples of $2 \pi$.

In general, however, Eq.s (\ref{gaucou98}) and (\ref{theta45}) are only
invariant under the $Z_2$ subgroup of $Z_{2N}$ corresponding
to the transformation:
\begin{equation}
\psi \rightarrow \psi + 2 \pi
\label{z2tra}
\end{equation}
that changes $\theta_{YM}$ in Eq.(\ref{theta45}) as follows
\begin{equation}
\theta_{YM} \rightarrow \theta_{YM} - 2 N \pi
\label{thtra}
\end{equation}
leaving invariant the gaugino condensate:
\begin{equation}
<\lambda^2 > = \frac{\mu^3}{3N g_{YM}^{2}} {e}^{- \frac{8 \pi^2}{N
    g_{YM}^{2}}}~~ {e}^{i \theta_{YM} /N}
\label{lala}
\end{equation}
Therefore the chiral anomaly and the breaking of $Z_{2N}$ to $Z_2$ are
encoded in Eq.s (\ref{gaucou98}) and (\ref{theta45}). Finally, if we
put $\psi=0$ in eq. (\ref{gaucou98}) and we consider it together with 
eq.(\ref{gauar78}) we can determine the running coupling constant as a
function of $\mu$:
\begin{equation}
\frac{4 \pi^2}{N g^{2}_{YM}} = \rho \coth 2 \rho - \frac{1}{2} a (
\rho) = \rho \tanh \rho
\label{gau84}
\end{equation} 
This equation taken together with eq.(\ref{gauar78}) reproduces~\cite{HOUCHE} 
the NSVZ $\beta$-function plus non-perturbative
corrections due to fractional instantons.


\subsection{The conifold solution}

In this second subsection we will start presenting the classical solutions of
type IIB supergravity corresponding to have $N$ fractional D branes
located at the tip of the conifold  and $M$ bulk D branes respectively
for the singular and the deformed conifold. Then we will use them to
get information on ${\cal{N}}=1$ super Yang-Mills.

The conifold is a manifold described by the following equation between
complex variables:
\begin{equation}
x^2 + y^2 + z^2 + t^2 = 0
\label{coni62}
\end{equation}
that can be seen as the six-dimensional cone over the space $T^{1,1}$,
so that the metric can be written as
\begin{equation}
ds_{6}^{2} = dr^2 + r^2 ds_{T^{1,1}}^{2} 
\label{ds6}
\end{equation}
where
\begin{equation}
ds_{T^{1,1}}^{2} = \frac{1}{9} (g^5 )^2 + \frac{1}{6} \sum_{i=1}^{4}
(g^i )^2
\label{t11}
\end{equation}
and
\begin{equation}
g^1 = \frac{e^1-e^3}{\sqrt{2}}~.~g^2 = \frac{e^2-e^4}{\sqrt{2}}~,~
g^3 = \frac{e^1+e^3}{\sqrt{2}}~,~g^4 = \frac{e^2+e^4}{\sqrt{2}}~,~
g^5 = e^5
\label{gs2}
\end{equation}
with
\begin{eqnarray}
e^1 = -\sin\theta_1 d\phi_1~,~e^2 = d\theta_1~,~
e^3 = \cos\psi \sin\theta_2 d\phi_2 - \sin\psi d\theta_2 \nonumber\\
e^4 = \sin\psi \sin\theta_2 d\phi_2 + \cos\psi d\theta_2~,~
e^5 = d\psi + \cos\theta_1 d\phi_1 + \cos\theta_2 d\phi_2
\label{es2}
\end{eqnarray}
The range of the angular coordinates is defined to be:
\begin{equation}
        0\leq\psi\leq 4\pi\,,\qquad
        0\leq\theta_1,\theta_2\leq \pi\,,\qquad
        0\leq\phi_1,\phi_2\leq 2\pi\,.
\label{range}
\end{equation}
Topologically the $T^{1,1}$ manifold can be thought of as $S^2 \times
S^3$. The two cycles are identified by:
\begin{eqnarray}
S^2 ~~ & : & \psi =0,~~\theta_1 = \theta_2,~~\phi_1 = -
\phi_2,\nonumber\\
S^3 ~~ & : & \theta_1 = \phi_1 =0
\label{s2s3}
\end{eqnarray}
Their volume forms are given respectively by:
\begin{equation} 
\omega_3 = \frac{1}{2} g^5 \wedge \left( g^1 \wedge g^2 + g^3 \wedge
  g^4 \right)   ~~,~~\omega_2 = \frac{1}{2} \left( g^1 \wedge g^2 + g^3 \wedge
  g^4 \right)
\label{volform}
\end{equation}
normalized as follows:
\begin{equation}
\int_{S^2} \omega_2 = 4\pi~~~,~~~\int_{S^3} \omega_3 = 8 \pi^2
\label{nor98}
\end{equation}
They satisfy the following duality relation in the six-dimensional
space transverse to the world-volume of a D3 brane: 
\begin{equation}
{}^{*6}\left( \omega_2 \wedge \frac{dr}{r} \right) =
\frac{\omega_3}{3}~~,~~
{}^{*6} \omega_3 = - 3 \omega_2 \wedge \frac{dr}{r}
\label{dua56}
\end{equation}
The classical solution corresponding to $N$ fractional at the tip of
the conifold and $M$ bulk branes is given by~\cite{KT}:
\begin{eqnarray}
ds^2  =  h^{-1/2} (r) \eta_{\alpha \beta}dx^{\alpha} d x^{\beta} + 
h^{1/2} (r) \left(dr^2 + r^2 ds_{T^{1,1}}^{2} \right) \label{ds}\\
B_2  =  \frac{3 g_s N \alpha'}{2} \log \frac{r}{r_0} \omega_2~~,~~
H_3 \equiv d B_2 = \frac{3 g_s N \alpha'}{2} \frac{dr}{r} \wedge \omega_2
\label{b2h3} \\
F_3 = \frac{ g_s N \alpha'}{2} \omega_3 
\label{f3}\\
{\tilde{F}}_5 = {\cal{F}}_5 + {}^* {\cal{F}}_5 ~~~.~~~{\cal{F}}_5 = 27 \pi
g_s (\alpha ')^2 M_{eff}(r) Vol ( T^{1,1} )~, \label{f5}
\end{eqnarray}
where $r_0$ is a regulator,
\begin{equation}
h (r) = 27 \pi (\alpha')^2 \frac{g_s M + \frac{3 (g_s N)^2}{2\pi}
  \left( \log \frac{r}{r_0} + \frac{1}{4}\right)}{4 r^2}~~,
\label{hr}
\end{equation}
\begin{equation}
M_{eff} (r) = M + \frac{3 g_s N^2}{2 \pi }\log \frac{r}{r_0}
\label{meff}
\end{equation}
The previous solution, corresponding to the singular conifold
described by eq. (\ref{coni62}), has a naked singularity at short
distances when
$r= r_0$. To remove this singularity the equation that defines the conifold is
replaced by:
\begin{equation}
x^2 + y^2 + z^2 + t^2 = \epsilon^2
\label{defo8}
\end{equation}
This corresponds to blow up the three-sphere $S^3$ of $T^{(1,1)}$. The
metric of the deformed conifold is given by~\cite{CANDE,TSI,OHTA}:
\begin{equation}
ds_{6}^{2} = \frac{\epsilon^{4/3}}{2} K(\tau) \left[ \frac{d \tau^2 +
(g^5 )^2}{3K^{3} ( \tau)}  + \cosh^2 \frac{\tau}{2} \left( (g^3 )^2 +
(g^4 )^2 \right) + \sinh^2 \frac{\tau}{2} 
\left( (g^1 )^2 + ( g^2 )^2 \right)\right]
\label{ds6de}
\end{equation}
where
\begin{equation}
K (\tau ) = \frac{(\sinh 2 \tau - 2 \tau)^{1/3}}{2^{1/3} \sinh \tau}
\label{ktau}
\end{equation}
The complete solution is given by~\cite{KS}:
\begin{eqnarray}
ds^2 =  h^{-1/2} ( \tau ) \eta_{\alpha \beta}dx^{\alpha} d x^{\beta} + 
h^{1/2} (\tau ) ds_{6}^{2} \label{ds31}\\
B_2 = \frac{g_s N \alpha'}{2} \left[ f (\tau) ( g^1 \wedge g^2 ) 
+ k (\tau) (  g^3 \wedge g^4 )   \right] \label{b21}\\
F_3 = \frac{ g_s N \alpha'}{2} \left[ g^5 \wedge  g^3 \wedge g^4 +
 d \left( F (\tau) ( g^1 \wedge g^3 +  g^2 \wedge g^4 \right) \right]
 \label{f31}\\
{\tilde{F}}_5 = {\cal{F}}_5 + {}^* {\cal{F}}_5 ~~~.~~~{\cal{F}}_5 = 
\frac{g_s N^2 (\alpha ')^2}{4} {\ell} (\tau ) g^1 \wedge g^2 \wedge  
g^3 \wedge  g^4 \wedge  g^5 ~, \label{f51}
\end{eqnarray}
where
\begin{equation}
F (\tau) = \frac{\sinh \tau - \tau}{2 \sinh \tau}~,~ 
f (\tau) =\frac{\tau \coth \tau -1 }{2 \sinh \tau} ( \cosh \tau -1)
\label{Ftau21}
\end{equation}
and
\begin{equation}
k(\tau ) = \frac{\tau \coth \tau -1 }{2 \sinh \tau} ( \cosh \tau +1)~,~
\ell (\tau ) = \frac{\tau \coth \tau -1 }{4 \sinh^2 \tau} 
(\sinh 2 \tau -2 \tau )
\label{ktau23}
\end{equation}
\begin{equation}
h (\tau ) = (g_s N \alpha' )^2 2^{2/3}
\epsilon^{-8/3} \int_{\tau}^{\infty} dx \frac{ x \coth x -1 }{ \sinh^2
  x} (\sinh 2 x -2 x )^{1/3}
\label{htau21}
\end{equation}
For large values of $\tau$  the previous regular solution
behaves as the singular one. In particular by identifying
eq. (\ref{ds6de}) with the six-dimensional part of eq. (\ref{ds})
without the warp factor we get a relation between $\tau$ and $r$ given
by:
\begin{equation}
\tau = 3 \log \frac{r}{r_0}~~~~,~~~r_0 \equiv 
\frac{3^{1/2} \epsilon^{2/3}}{2^{5/6}}
\label{rtau7}
\end{equation}
Let us now use the previous solution for obtaining the gauge couplings
of ${\cal{N}}=1$ super Yang-Mills given in eq.s (\ref{volu87}) and
(\ref{theta56}). In the case of the conifold eq.s (\ref{volu87}) and 
(\ref{theta56}) become:
\begin{equation}
\frac{4\pi}{g_{YM}^{2}} = \frac{1}{g_s ( 2 \pi \sqrt{\alpha'})^2}
\int_{S^2} B_2~~,~~ \theta_{YM}= \frac{1}{2 \pi \alpha' g_s}
\int_{S^2} C_2
\label{coupli92}
\end{equation} 
We start by computing the
$\theta_{YM}$ parameter. We need to extract $C_2$ from
eq. (\ref{f31}). It is given by:
\[
C_2 = \frac{N g_s \alpha'}{4} \left[(\psi + \psi_0 ) 
\left( \sin \theta_2 d \phi_2
  \wedge d \theta_2 - \sin \theta_1 d \phi_1 \wedge d \theta_1\right) +
\right.
\]
\[
-  d\phi_1 \wedge d \phi_2
  \cos \theta_1 \cos \theta_2  + ( 1 - 2 F(\tau)) 
\left( \sin \psi ( \sin \theta_2 d \phi_2 \wedge d
    \theta_1 - \sin \theta_1 d \phi_1 \wedge d \theta_2) + \right.
\]
\begin{equation}
\left. + \cos
    \psi \left( d\theta_1 \wedge d \theta_2 - \sin \theta_1 \sin
    \theta_2 d \phi_1 \wedge d \phi_2 \right)  )\right] 
\label{c289}
\end{equation}
When we insert $C_2$ in the second equation in (\ref{coupli92}) and we
take $\theta_1 = \theta_2, \phi_1 =- \phi_2$ we get
\begin{equation}
\theta_{YM} = N \left[ ( \psi + \psi_0 ) + \sin \psi ( 1 - 2 F (\tau))
  \right]
\label{theta57}
\end{equation}
showing as in the case of the Maldacena-N{\'{u}}{\~{n}}ez solution
that, for large values of $\tau$ where $F(\tau) = \frac{1}{2}$ the
$U(1)_R$ is broken down to $ Z_{2N}$. This is also the symmetry of the
asymptotic solution, while the full solution presented above does not
have anymore this $Z_{2N}$ symmetry. It is further broken to $Z_2$
that is the remaining symmetry of ${\cal{N}}=1$ super Yang-Mills that
leaves invariant the gaugino condensate. This observation gives us the
possibility of again identifying the gluino condensate. The natural thing is
to identify it with $2 F(\tau)$ after having subtracted its asymptotic
value~\cite{EI}:
\begin{equation}
1  - 2 F (\tau) = \frac{\tau}{ \sinh \tau }
\label{gaugi4}
\end{equation}
This quantity will play the same role as $a(\rho)$ in eq. (\ref{equa})
and allows one to establish the relation between the supergravity
parameter $\tau$ and the renormalization group scale $\mu$:
\begin{equation}
\frac{\Lambda^3}{\mu^3} = \frac{\tau}{ \sinh \tau }
\label{gaugra3}
\end{equation}
Notice the strong similarity between this equation and
eq. (\ref{gauar78}) and also between eq.s (\ref{theta45}) and (\ref{theta57}).
The running coupling constant can then be computed by inserting
eq.(\ref{b21}) in the first equation in (\ref{coupli92}) getting:
\begin{equation}
\frac{4 \pi^2}{g^{2}_{YM}} =  N  k(\tau ) = N \frac{\tau
  \coth \tau -1 }{2 \sinh \tau} ( \cosh \tau +1) 
\label{run85}
\end{equation}
that taken together with eq.(\ref{gaugra3}) allows one to compute the
$\beta$-function of ${\cal{N}}=1$ super Yang-Mills. A simple
calculation shows that
\begin{equation}
\beta ( g_{YM} ) \equiv \mu \frac{\partial g_{YM}}{\partial \mu} = -
\frac{3N g^{3}_{YM}}{16 \pi^2} \cdot \frac{1 + \frac{\tau}{\sinh
    \tau}}{\coth \tau - \frac{1}{\tau}}
\label{beta87}
\end{equation}
where $\tau$ is a function of $g_{YM}$ given by eq.(\ref{run85}). In
the ultraviolet ($ \tau \rightarrow \infty$) one can get a more 
explicit relation between them given by:
\begin{equation}
\frac{1}{\tau} = \frac{ \frac{N g_{YM}^{2}}{8 \pi^2} }{1 +\frac{N
    g_{YM}^{2}}{8 \pi^2} } \cdot \left[  1 + \frac{ {e}^{- 8 \pi^2 /(N
    g_{YM}^{2})}}{1 +\frac{N
    g_{YM}^{2}}{8 \pi^2}} \right]
\label{rela87}
\end{equation}
Inserting it in eq.(\ref{beta87}) and neglecting the contribution of
the fractional instantons we get:
\begin{equation}
\beta ( g_{YM} ) = -  \frac{3N g^{3}_{YM}}{16 \pi^2} \cdot \frac{1}{ 1 -
  \frac{Ng_{YM}^{2} }{8 \pi^2}[1 + \frac{Ng_{YM}^{2}}{8 \pi^2}]^{-1}  }
\label{beta88}
\end{equation}
that agrees with the NSVZ
$\beta$-function up to two loops, but differs from it for higher
loops. Notice that, if one takes into account only the leading
asymptotic behaviour of the right hand side of eq.(\ref{run85}), one
gets exactly the NSVZ $\beta$-function~\cite{EI}.

In the case of the Klebanov-Strassler solution one can also compute
the effective potential of ${\cal{N}}=1$ super Yang-Mills, namely the 
Veneziano-Yankielowicz potential, following a proposal by
Vafa~\cite{VAFA} where
he identifies it with the superpotential of ${\cal{N}}=1$ supergravity
that is given by the following expression:
\begin{equation}
W_{eff}  \sim \sum_{i} \left[ \int_{A_i} G_3 \int_{B_i} \Omega - 
\int_{A_i} \Omega \int_{B_i} G_3 \right]
\label{supo67}
\end{equation} 
in terms of the flux of $G_3 = F_3 + i H_3$ and of the periods of the 
holomorphic
$(3,0)$-form $\Omega$ of Calabi-Yau threefold under consideration.
$A_i ( B_i )$ are the compact (noncompact) orthogonal
three-cycles of a Calabi-Yau manifold.

In the following we will use the previous general formula in the case
of the deformed conifold solution and we will obtain~\cite{VI} the
Veneziano-Yankielowicz potential~\footnote{The procedure followed here
  is taken from Ref.~\cite{EMILI}.}. In the case of the
conifold we have
only one compact and one non-compact cycle specified by 
\begin{eqnarray}
A (compact)~~:~~~~  r= constant, \theta_1 =0, \phi_1 =0 \\
\label{com9}
B (noncompact)~~,~~~ \psi =0, \theta_1 = \theta_2, \phi_1 = - \phi_2
\label{noncompa}
\end{eqnarray} 
Let us start computing the fluxes of $G_3$ along the two cycles. For
both the singular and deformed conifold solution the flux of $G_3$ on
the compact cycle is given by:
\begin{equation}
\frac{1}{(2 \pi \sqrt{\alpha'})^2 g_s} \int_{A} G_3 = N
\label{fluxa}
\end{equation}
In the case of the non compact cycle in order to get finite
expressions we need to introduce a cut-off $r_c$ for large values of
$r$. In addition in the case of the singular solution we also need to
introduce a cut-off $r_0$ for small values of $r$. In the case of the
singular solution we get:
\begin{equation}
\int_B F_3 =0~~~,~~~ \int_B H_3 = 6 \pi g_s N \alpha' \int_{r_0}^{r_c}
\frac{dr}{r} = 6 \pi g_s N \alpha' \log \frac{r_c}{r_0}
\label{singu78}
\end{equation}
while in the case of the deformed we get:
\begin{equation}
\int_B F_3 =0~~~,~~~ \int_B H_3 = 4 \pi g_s N \alpha'
\int_{0}^{\tau_c}  d \tau k' (\tau) =4 \pi g_s N
\alpha' k (\tau_c ) 
\label{defo78}
\end{equation}
that implies
\begin{equation}
\frac{1}{(2 \pi \sqrt{\alpha'} )^2 g_s} \int_B G_3 = \frac{2N k(
  \tau_c )}{2 \pi i} \sim \frac{2N 
  \tau_c }{4 \pi i} = \frac{3N}{2 \pi i} \log \frac{r_c}{r_0}
\label{g3b}
\end{equation}
where we have used eq.(\ref{rtau7}) and the behaviour of $k(\tau_c)
\sim \frac{\tau_c}{2}$ for large $\tau_c$.

Let us now compute the two periods of the holomorphic $(3,0)$-form
$\Omega$. The deformed conifold is described by the equation:
\begin{equation}
F = x^2 + y^2 + z^2 + t^2 - \epsilon^2 =0
\label{Fconi}
\end{equation} 
and $\Omega$ is defined as
\begin{equation}
\Omega = \frac{1}{2 \pi i} \oint_{F=0} \frac{dx \wedge dy \wedge dz
  \wedge dt}{F}= 
\frac{dx \wedge dy \wedge dz}{2 \sqrt{\epsilon^2 - x^2 -y^2 - z^2}}
\label{OMEGA1}
\end{equation}
The cycle A is determined by letting  $x$ and $y$ to run
in the intervals $-\epsilon \leq x \leq \epsilon; - \sqrt{\epsilon^2
  -x^2} \leq y \leq \sqrt{\epsilon^2
  -x^2}$ and $z$ around the
 branch cut that connects the two branch points $\pm
\sqrt{\epsilon^2 - x^2 - y^2}$. We have to compute therefore the
following expression:
\begin{equation}
\int_{A} \Omega = \int_{- \epsilon}^{\epsilon} dx \int_{- \sqrt{\epsilon^2
  -x^2}}^{\sqrt{\epsilon^2
  -x^2}} dy \int_{\gamma} \frac{dz}{2\sqrt{\epsilon^2 - x^2 -y^2 - z^2} }
\label{OMEGA2}
\end{equation}
where $\gamma$ is a curve around the branch cut. The integral over $z$
can be computed by deforming it to an integral around infinity and one
gets:
\begin{equation}
\int_{A} \Omega = \int_{- \epsilon}^{\epsilon} dx \int_{- \sqrt{\epsilon^2
  -x^2} }^{\sqrt{\epsilon^2
  -x^2}  }
dy \oint_{\infty} \frac{dz}{2i z} =\int_{- \epsilon}^{\epsilon} dx
 2 \pi  \sqrt{\epsilon^2 -x^2} = \pi^2 \epsilon^2
\label{peri89}
\end{equation} 
In the case of the cycle B we get instead:
\[
\int_{B} \Omega = \int_{\epsilon}^{r_{c}^{3/2}} dx 2\pi
\sqrt{\epsilon^2 -x^2} = 2 \pi \epsilon^2 \int_{\pi/2}^{arcsin
  \frac{r_{c}^{3/2}}{\epsilon} }  d \alpha \cos^2 \alpha= 
\]
\begin{equation}
= \pi r_{c}^{3/2} \sqrt{\epsilon^2 -r_{c}^3} + \pi \epsilon^2 \arcsin 
\frac{r_{c}^{3/2}}{\epsilon} - \frac{1}{2} \pi^2 \epsilon^2
\label{perib}
\end{equation}
where we have taken care that the complex coordinates in
eq.(\ref{Fconi}) have dimension $L^{3/2}$ as you can see from 
eq.(\ref{rtau7}). Expanding eq. (\ref{perib}) for large values of
$r_c$ we get:
\begin{equation}
\int_{B} \Omega = 2 \pi i \left[ \frac{r_{c}^{3}}{2} -
  \frac{\epsilon^2}{4} +\frac{\epsilon^2}{4} \log \frac{\epsilon^2}{4}
  - \frac{\epsilon^2}{2} \log r_{c}^{3/2}\right]
\label{perib2}
\end{equation} 
Putting together eq.s (\ref{fluxa}),  (\ref{g3b}),  (\ref{peri89}) and  
(\ref{perib2}) we get the following effective potential for
${\cal{N}}=1$ super Yang-Mills:
\[
W_{eff} = - \frac{1}{2 \pi i} \frac{1}{( 2 \pi \sqrt{\alpha'})^2 g_s }
\frac{1}{(2 \pi \alpha')^3 } \left[ \int_{A} G_3 \int_{B} \Omega - 
\int_{A} \Omega \int_{B} G_3 \right]=
\]
\begin{equation}
= - \frac{N}{( 2 \pi \alpha ')^3} \left[ 3 \frac{\epsilon^2}{4} \log
    \frac{r_c}{r_0 } + \frac{r_{c}^{3}}{2} - \frac{\epsilon^2}{4} + 
\frac{\epsilon^2}{4} \log \frac{\epsilon^2}{4r_{c}^{3}} \right]
\label{weff}
\end{equation}
Making the following identifications:
\begin{equation}
r_c = 2 \pi \alpha' \mu~~,~~r_0 = 2 \pi \alpha'
\Lambda~~,~~\frac{\epsilon^2}{4} = (2 \pi \alpha')^3 S 
\label{ide34}
\end{equation}
and neglecting the constant term
we get the Veneziano-Yankielowicz effective superpotential:
\begin{equation}
W_{eff} = N S \left( 1 - \log \frac{S}{\Lambda^3}\right)
\label{vy}
\end{equation}
Finally let me mention that the previous procedure for computing the
effective superpotential has been also used for computing~\cite{IL} the
Affleck-Dine-Seiberg superpotential in the case of the orbifold $C^3
/(Z_2 \times Z_2)$. 

\subsection*{Acknowledgements}

I thank E. Imeroni for explaining me many details of his PhD thesis
and A. Lerda and P. Merlatti for discussions.

\end{document}